# Adhesive Shear Strength of Ice from Nanostructured Graphite Surfaces by Molecular Dynamics Simulations

Amir Afshar,[*] and Dong Meng [§]
Dave C. Swalm School of Chemical Engineering, Bagley College of Engineering, Mississippi State University, Starkville, MS, 39762, USA

(Dated: 22 November 2020)

The issue of ice accumulation at low-temperature circumstances causes multiple problems and serious damages in many civil infrastructures which substantially influence human's daily life. However, despite the significant consideration in manufacturing anti-icing or icephobic surfaces, it is still demanding to design surfaces with well ice-repellent properties. Here in this study, we used all-atom molecular dynamics (MD) simulations to investigate ice shearing mechanism on atomistically smooth and nanotexture graphite substrates. We find that ice shearing strength strongly depends on ice temperature, the lattice structure of the surface substrate, the size of the surface nanotexture structure, and the depth of interdigitated water molecules. Our results indicate nanoscale surface roughness and depth of interdigitated water molecules tend to increase ice shear failure stress and for corrugated substrates, this is further raised with increasing the depth of interdigitated water molecules which is a result of strain being distributed well into the ice cube away from the interface. These results supply an in-depth understanding of the effect of surface nanotexture on ice shearing mechanism that provides useful information in designing anti-icing surfaces and provide for the first-time theoretical references in understanding the effect of surface nanotexture structure and depth of interlocked water on adhesive ice shear strength on nanotextured surfaces.

## I. INTRODUCTION

Ice accumulation on solid surfaces plays a considerable role in many industries. It can afford a significant serious issue in the reduction of operational safety of offshore and performance of solar panels and wind turbines[1], [2] In some cases, excessive ice accumulation results in a crash of airplanes due to increasing the drag force exerted on aircraft [3], intensive damage to infrastructure including building, transmission lines, bridges, off-shore oil instruments which eventually leads to economic disadvantages[4]–[6]. Up to now, great research attempts have been devoted to recognizing the mechanism of icing which helps to extend the impressive anti-icing method[7]–[9]. One of the widely used methods is classified as active anti-icing methods, which are based on mechanical removal of the ice layer, surface chemical modifications, and heating treatments. These traditional methods are not only cost-ineffective and environmentally unfriendly, but incapable of surface protection from ice-build-up for a longer duration of exposure[10]–[12]. Thus, the difficulty of de-icing has inspired fundamental anti-icing research study through shedding light on recognizing the basics of ice adhesion[13], [14] to determine the major interactions responsible for ice adhesion strength. Several recent research studies have been devoted to relating ice adhesion to the wettability of surface substrate[15], [16].

---

[*] Email: aa2552@msstate.edu
[§] Email: meng@che.msstate.edu

This resulted in the effort of applying a hydrophobic surface coating for anti-icing applications[17]–[20]. Graphene is the most common hydrophobic coating material which is mainly used in aerospace industries. Consisting of one-atom-thick sheets of linked carbon atoms in a hexagonal lattice structure, graphene layers are highly thermally and electrically conductive which makes the airplane wings to be resistant to ice buildup[21]. Also, it has been newly demonstrated that a thin ( ~nm) disordered layer of interfacial water molecules between coated surface and ice in experiments gives rise to an increase in the deicing performance of surface through decreasing ice adhesion strength[22]–[24]. Although, access to the nanoscale mechanics through experimental measurements is limited by their temporal and spatial resolutions. Atomistic level understanding of ice adhesion mechanics on nanostructured surfaces is still limited and expose to promising arguments.

Since it is believed that the delamination of ice from the substrate is likely a combination of shear and normal modes of ice failure mechanism, here we describe straightforward computational modeling and molecular dynamics simulations proof of ice adhesion on nanostructured graphite surfaces through shedding light on only shear simulation of ice cubes using the atomistic model of water. In this study, we initially construct a nano-sized ice cube model on various graphite surfaces indicating different surface texture sizes. We, then equilibrate our setup models by relaxing the interfacial ice layer on different graphite substrates to obtain an established ice cube adhesion state. Finally, we load shear forces, parallel with respect to ice-substrate contact area, on ice cubes to investigate shear stresses of ice from different nanostructured graphite substrate. The findings of this study bring us a step forward to identifying the relevant physical parameters that control ice-substrate shearing strength. The following developed simulation protocols give clearance in further systematic examinations on ice fracture at various types of ice-substrate interfaces.

## II. MODELS AND METHODS

Here we describe our modeling and simulation strategy. Our general objective is to study the failure dynamics of a nanoscale ice cube on nanostructured graphite surfaces through performing force-probe shear simulations.

### A. MODEL SETUP

In this study, all-atom TIP4P/Ice[25], [26] force field is used to model water molecules by using the interaction parameters shown in Table 1. Comparing to the other widely used atomistic water models, such as SPC[27], TIP3P[28], and TIP4P[29], this water model was described to reproduce ice and water physical properties and phase transition at low temperatures. It has a higher transition temperature of 269.8 ±0.1 K[30] which is favorable compared to its corresponding experimental value of 273.15K, thus it is appropriate in this work for examining shear failure strength of ice on solid substrates. We choose to model ice $I_h$ having a hexagonal arrangement, as shown in Figure 1, with the initial coordinates taken from Matsumoto et. al[31]. Ice cube has a surface area of A = 6.6 × 9.1 nm$^2$ with a thickness of 4.7 nm normal to the ice-surface interface contains primary prismatic plane {10$\bar{1}$0} with 8640 number of water molecules. This face will be used to adhere to the substrates.

Table 1. Interaction parameters for the four-site water model, TIP4P/Ice[25]

| Water model | $\varepsilon_{O-O}$ [kj.mol$^{-1}$] | $\sigma_{O-O}$ [Å] | $q_H$ (e) | $q_O$ (e) | $d_{OM}$ [Å] |
|---|---|---|---|---|---|
| **TIP4P/Ice** | 0.8822 | 3.1668 | 0.5897 | -1.1794 | 0.1577 |

In Table 1, $q_H$ and $q_O$ are the electronic charges carried by hydrogen and oxygen atoms respectively and $d_{OM}$ is the distance between oxygen atoms and massless charged points in the water model. A graphite substrate consisting of multiple layers of zig-zag edge graphene sheets with an inter-sheet distance of 3.35Å[32] is considered as the substrate material in this study[33]. Each graphene sheet consists of a hexagonal arrangement of 1296 carbon atoms which are bonded to their closest neighbors at the equilibrium bond length and bond angle of 0.142Å and 120° respectively [33]–[35]. For simplicity, atoms of graphite substrates in the present study are electronically neutral and do not interact with each other throughout the simulations. Similar conjectures were also made in the previous ice fracture studies[11], [12], [13] and wetting studies[35], [39]–[41] using MD simulations. It has been reported that electrostatic interactions and van der Waals forces are both responsible for ice adhesion on surfaces[13], [42]. Because of the probable surface oxidation event caused by coulombic forces of the substrate on ice[36], we here confine ourselves to only examine van der Waals interactions, i.e. the Lennard Jones (LJ) potential, between water molecules of ice structure and substrate carbon atoms.

Note that the atomistic interaction parameters of LJ potential between oxygen atoms of water molecules and carbon atoms of graphene sheets vary in the published literature. On the one hand, following the previous study[43], we select oxygen-carbon LJ potential parameters as $\varepsilon_{co}$ = 0.4736 kJ/mol and $\sigma_{co}$ = 3.19Å that can reproduce experimentally observed water contact angle (WCA) on a clean graphite surface[44]. On the other hand, we varied LJ potential parameters as $\varepsilon_{co}$ = 0.6887 kJ/mol and $\sigma_{co}$ = 3.126Å indicating WCA lower than 20°[45]. These two different sets of interaction parameters are used for ice-substrate system equilibration.

We initially placed a pre-equilibrated ice cube into the nearest possible distance with atomistically flat and nanotextured graphite surfaces by the energy minimization process. Flat graphite substrates consist of graphene sheets oriented parallel (PA- the snapshot on the left) and perpendicular (PE-the snapshot on the right) with respect to the ice-surface interface as shown in Figure 1(a). Textured substrates are modeled by constructing two graphite blocks made of PE cases with introducing nanoscale roughness which is separated by a distance *d*, as illustrated in Figure 1(b). Graphite substrates have a similar contact area with the ice cube along the lateral dimension, i.e. yz plane with the thickness of 5nm perpendicular to the ice-surface contact area. Because of the periodic boundary condition (PBC) applied in all principle directions, the ice-substrate interface is bulk-like in our modeled setups. This marks one important feature and difference of our study with the other published literature[36], [37], [46]–[48].

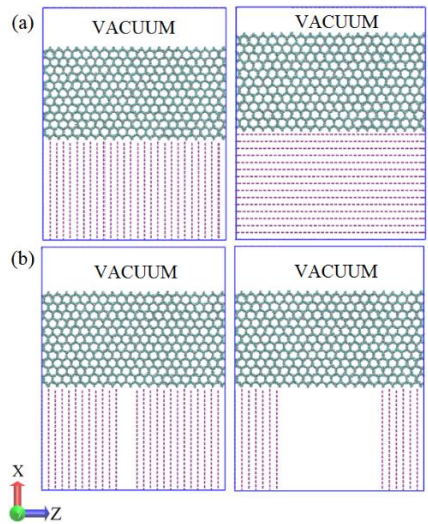

**FIG. 1.** Snapshots of modeled systems consist of pre-equilibrated ice $I_h$ on different flat graphite substrates contain (a) PE (on the left) and PA (on the right) cases, and (b) texture graphite substrates with nano roughness size of d=1 nm (on the left) and d=5 nm (on the right).

## B. SIMULATION METHODOLOGY

To investigate the shear failure stress of ice on different graphite substrates, we performed two different subsequent equilibrium and non-equilibrium MD simulations on modeled setups namely as ice-substrate equilibration and force-probe MD shear simulations, respectively[49]. All simulations are performed using the open-source MD code, LAMMPS[50] simulation package. To decrease unrealistic interactions between mirror images at any time during the simulation, the X-dimension of the simulation box is increased by two times larger than the cutoff distance for non-bonded LJ interactions (1.0 nm) [33], [35], [38]. Thus, the simulation box size is $12.0 \times 6.6 \times 9.1$ nm$^3$. The long-range electrostatic interactions for the ice/water molecules are calculated using the particle-particle particle-mesh (P$^3$M)[51] method and water molecules are treated as rigid bodies by constraining the bond length and bond angles using the SHAKE algorithm[52]. Unless otherwise noted, all simulations are performed at a temperature of 250K and we use the Nośe–Hoover coupling method to maintain the simulation temperature[53], [54], with a coupling time constant $\tau_T = 0.2$ps. The equations of motion were integrated with the Velocity-Verlet algorithm using a 0.002 ps timestep.

### (1) EQUILIBRIUM ICE-SUBSTRATE MD SIMULATION

To equilibrate the ice-substrate systems shown in Figure 1, we initially introduced a liquid-solid interface along the prismatic plane by melting 15Å thickness of interfacial ice layer at an elevated temperature of 300K. This functions as the nucleation sites in which liquid state water molecules turn into ice crystals[55]–[58] by quenching all water molecules to its initial equilibrium temperature. For the ice cube on textured graphite substrates, various thickness of interfacial ice layers is melted which enables us to access the effect of penetration depth (PD) of interlocked water molecules on failure dynamics of ice during shear simulations.

To monitor the extent of ice-substrate equilibration, a water molecule needs to be labeled as being in the "solid-state" or "liquid-state". For a given water molecule *i*, the Steinhardt order parameter[59], [60] can be calculated:

$$q_l(i) = \sqrt{\frac{4\pi}{2l+1} \sum_{m=-l}^{l} \left(q_{lm}(i)\right)^2} \qquad (1)$$

$$q_{lm}(i) = \frac{1}{N_b(i)} \sum_{j=1}^{N_b(i)} Y_{lm}(r_{ij}) \qquad (2)$$

where $N_b(i)$ is the number of neighboring water molecules, *l* is a free integer parameter, $Y_{lm}$ is the spherical harmonics function and $r_{ij}$ is the vector from water molecule i to water molecule j. In order parameter calculation, typically $q_4$ and $q_6$ are often used as they are a good choice to distinguish between cubic and hexagonal solid structures[61], [62]. Figure 2 shows $q_4$ and $q_6$ values obtained for a bulk of 8640 number of water molecules in the liquid state (red points) and solid ice $I_h$ (blue points) at 1 bar and 250K. Comparing to $q_4$ values, it is obvious $q_6$ alone is a better parameter to differentiate between solid-like and liquid-like water molecules[63]. To distinguish liquid water from solid ice structure, we set a threshold value of $q_6 = 0.37$[56], indicated by a horizontal dashed line in Fig. 2, to define a water molecule to be in the solid-state and "liquid-state" otherwise.

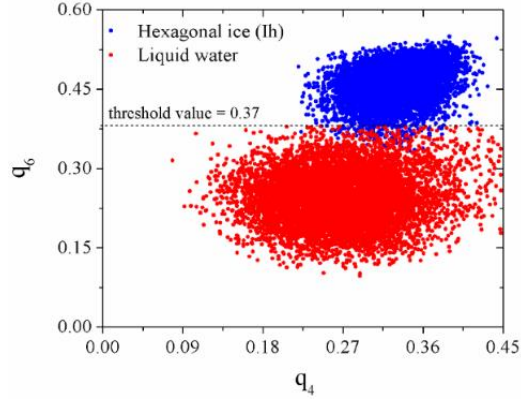

**FIG. 2.** Values of q6 and q4 for 8640 water molecules of the liquid water (red points), and ice-Ih (blue points) at 250K and 1 bar for the TIP4P/ice water model.

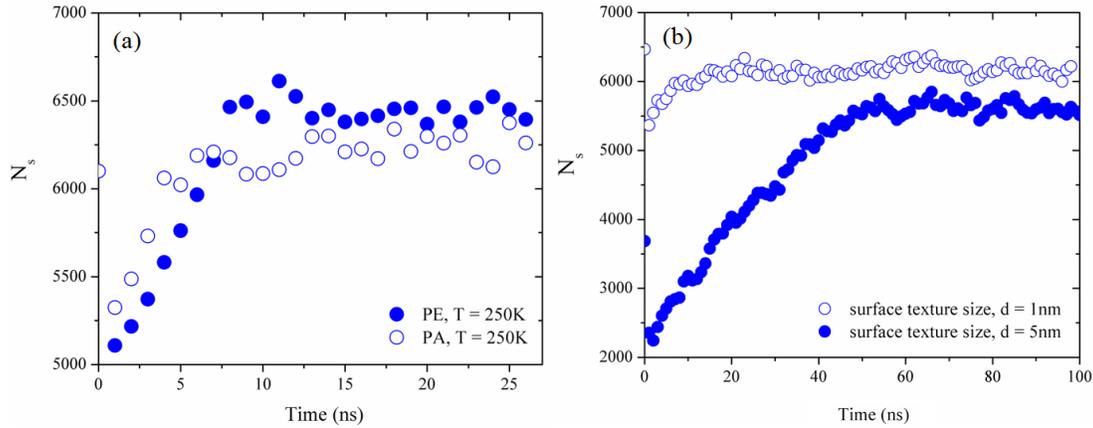

**FIG. 3.** Number of water molecules in solid-state (Ns) as a function of simulation time calculated during ice-substrate equilibration simulations at 250.0K for the systems of the ice cube on (a) flat graphite substrates and (b) graphite substrates with surface roughness sizes of 1nm, 2nm, 3nm, 4nm, and 5nm.

Figure 3 shows the evolution of the number of water molecules in solid-state ($N_s$) as a function of simulation time for the flat and textured graphite substrates at 250 K which is used as the indicator of the extent of interfacial equilibration. Figures show an initial abrupt drop in the number of $N_s$ values for the ice cube on different substrates that is due to the ice-vacuum and ice-supercooled liquid water interfaces. Following that, a significant increase in the number of $N_s$ values is observed during which ice crystallization from supercooled liquid water happens. Eventually, $N_s$ values demonstrate a developed plateau behavior representing equilibration of ice-substrate equilibration in which the corresponding equilibrated structures used as initial configuration for performing shear simulations. Figure 4 represents the systems of the ice cube in flat and textured graphite substrates that are fully equilibrated at 250K before simulations to measure shear failure stress are conducted. Notice the molecular disordering structures obvious at the ice-substrate interface in all three cases as the results of the mismatch between lattice structures of ice phase and graphene sheets.

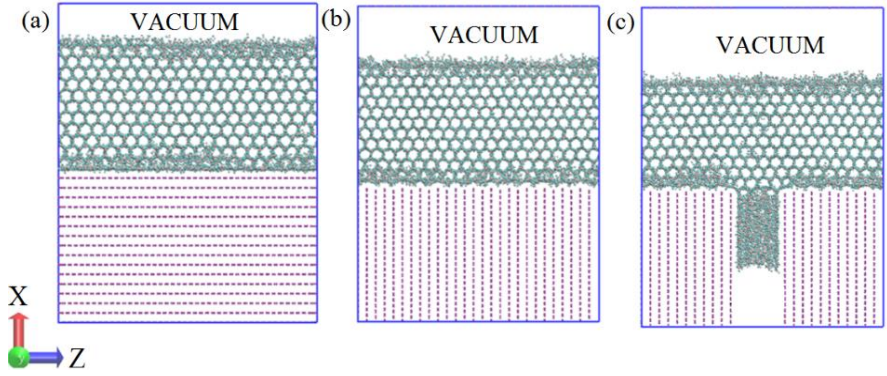

**FIG. 4.** Snapshots of equilibrated systems of ice on the: (a) PA, (b) PE, and (c) PE substrate with nanoscale roughness of d.

## (2) NON-EQUILIBRIUM MD SIMULATION

To investigate the shear failure strength of the ice cube on different flat and textured graphite substrates, we performed force-probe shear simulations by applying an external force on all water molecules located in the upper region (15Å from the ice-vacuum interface) of the ice cube along the parallel and perpendicular to the ice-substrate interface, i.e. y- or z-direction. The magnitude of the loading shear force starts at zero and increased at a constant rate of $3.7162 \times 10^{-5}$ per MD timestep. The trajectories, velocities, and forces corresponding to all the atoms in the system were saved every 2.0ps. During simulations, the interaction forces between oxygen atoms of the water molecules and carbon atoms of the graphene sheets are measured and recorded. The interfacial shear stress is then calculated by dividing the y- or z-component of the interfacial interaction forces by the ice-substrate contact area as:

$$\tau_{shear} = \frac{F_{shearing}}{ice - substrate\ contact\ area\ (A)} \quad (3)$$

Figure 5(a) shows an exemplified displacement of the ice cube experienced shear simulation along z-direction as a function of coordinates of all water molecules along the x-direction. The data indicate a high disparity in the domain of ~ x[0,8]Å corresponds to the interfacial region and ice shear strain is obtained by fitting the data using a linear function. The shear rate in these simulations is 0.13s$^{-1}$ that is determined by linear fitting the calculated shear strain as a function of simulation time as shown in Figure 5(b). Figure 5(c) illustrates an example of the measured interfacial shear stress as a function of the shearing distance for a PE substrate sample with d = 1.0 nm. Shearing distance is measured as the change in center of mass (COM) of the ice cube that water molecules experienced along the shearing direction. The interfacial shear stress shows a two-stage variation as a function of shearing distance. Through linear fittings of the first and second segments of data shown as red lines in Figure 5(c), the interfacial shear failure stress is determined as the point that two fitted lines are intersected as represented by the red point in Figure 5(c). To ensure a valid statistical representation and enhance the statistics of measurements, we conducted 10~50 shear simulations using different initial configurations which are generated one nanosecond apart during the interfacial equilibration stage. Figure 5(d) is a representative of the probability distribution of measured shear failure stress analyzed with a bin-size of 5.0Mb. We then fit the distribution using the Gaussian function indicated by the red line in which the mean and standard deviation of shear failure stress for a given system is deduced.

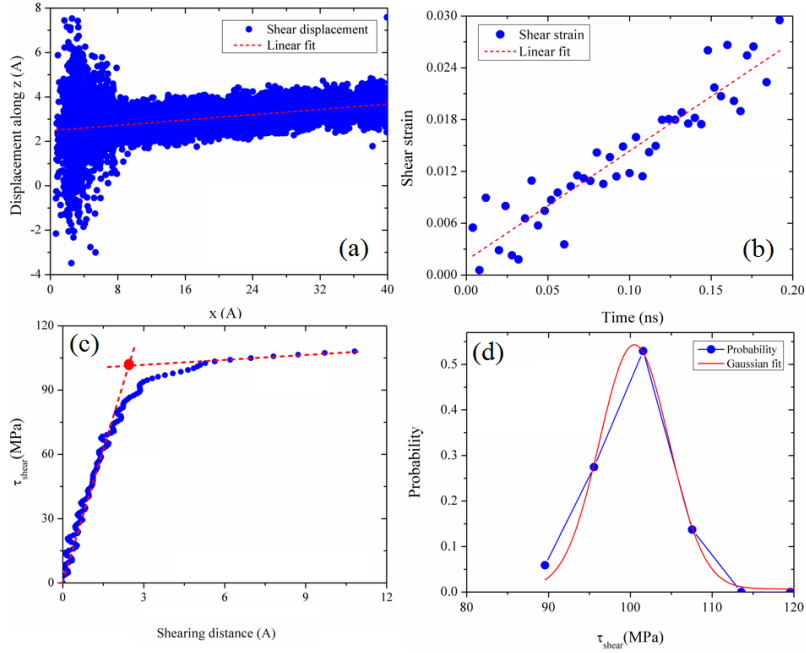

**Fig. 5**. (a) Examples of shear stress-strain curves from the simulation of ice on graphite substrates with $d = 1.0$ nm; (b) distributions of the measured shear strength magnitudes.

## III. RESULTS AND DISCUSSIONS

### A. STATIC STRUCTURE

To characterize the equilibrium structure of the ice cube on PA and PE graphite substrates, we calculate the distribution of the $q_6$ order parameter for all water molecules, Q6 (x), as a function of ice thickness, i.e. along x-direction shown in Figure 6(a) and (c). This parameter determines the "ordering" of the ice structure as a function of the distance from the substrate. At a given thickness of the ice cube along the x-direction, the Q6 (x) demonstrates greater value if the structure is more crystal-like. To distinguish between crystal-like and liquid-like structure within the ice cube, we set a threshold value indicated by a dashed line in Figure 6(a) and (c). The distributions of Q6 (x) indicate that except for the water molecules have the vacuum and substrate interfaces shown in Figure 6(b) and (d), the structure of water molecules is obviously crystal-like for the ice cubes on both PA and PE substrates at various four temperatures. This domain of ice cube is labeled as "ice block". Close to the vacuum region as well as the substrate material, results declare a remarkable deterioration in structure "ordering" with a thickness of ~10Å indicating liquid-like disorder structure, which is due to the mismatch between the lattice structure of the graphene sheets and the ice phase in the latter case.

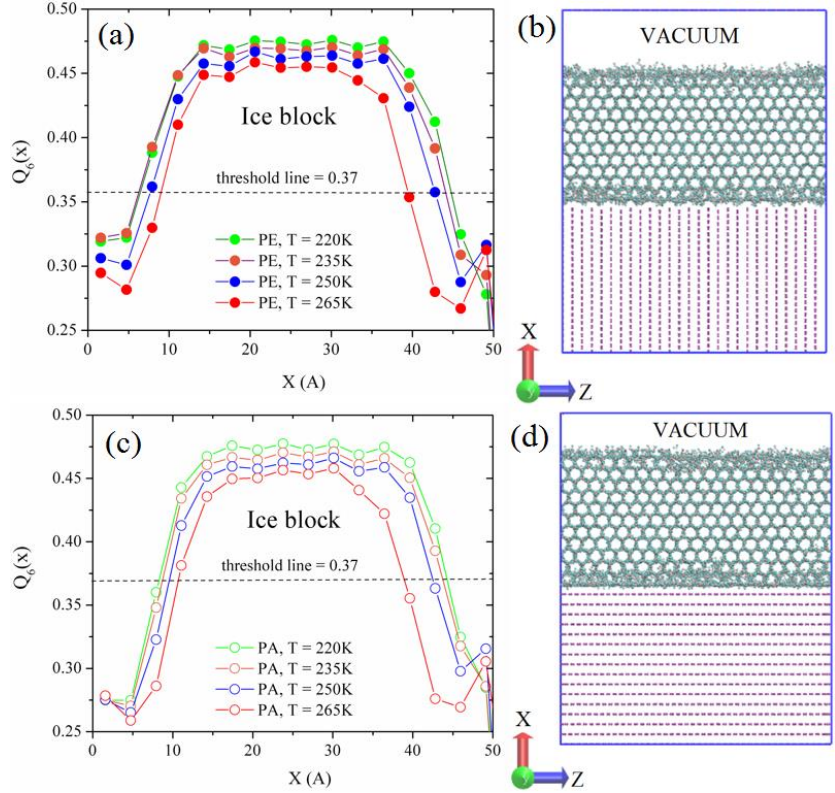

**Fig. 6.** (a) Q6(x), distribution of q6 order parameter along "x" direction normal to the ice-surface interface at different four temperatures of 220K, 235K, 250K, and 265K and (b) the corresponding snapshots of an equilibrated ice cube on PE substrates and (c-d) Q6(x) and the relating equilibrated system snapshot for ice cube on PA substrates.

### B. ICE SHEARING MECHANICS

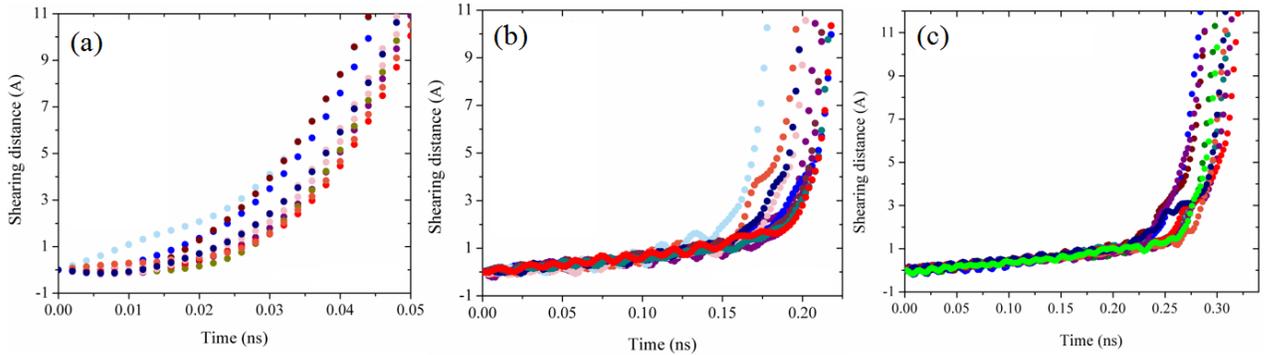

**FIG. 7.** Shearing distance of ice cube experienced during shear simulations along z-direction as a function of simulation time for the ice cube on (a) PA, (b) PE and (c) PE with nanoscale roughness size $d$ of 1.0 nm.

Figure 7 shows exemplified shearing profiles of ten equilibrated systems of an ice cube on PA and PE substrates as a function of simulation time at a temperature of 250 K. The shearing distance profiles exhibit a two-stage variation with simulation time in which the ice cubes demonstrate an acceleration by indicating a yield strength at a specific simulation time once loading shear force dominates the ice-substrate interacting forces. Furthermore, results indicate accelerations occur at higher simulation times for the ice cubes on PE compared to the PA substrates. This reflects the higher interfacial shear failure stress of the ice on PE substrates compared to PA cases as it is validated by the results shown in Figure 3. Also, shearing profiles of ice cubes on PE substrates (Fig. 7(b-c)) illustrate crooked shearing displacement profiles before demonstrating the acceleration compared to the smooth displacement profile obtained in PA substrate (fig 6(a)). We address this to the atomistically flat structure of the PA substrate compared with the PE cases.

### C. EFFECT OF SURFACE TEXTURE STRUCTURE AND ICE TEMPERATURE ON INTERFACIAL SHEAR STRENGTH OF ICE

Figure 8(a) illustrates the averaged shear failure stress of ice cube on PA and PE graphite substrates obtained from several independent shear simulations at various four temperatures of 220.0 K, 235.0 K, 250.0 K, 265.0 K. It shows ice cubes indicate much higher interfacial shear stress on PE substrates at different temperatures compared to the PA cases, which is reasonable given the mitigation behavior of ice cubes during the shear simulations shown in Figure 2. We address this difference to the distinction in the texture structure of the surface along which the ice cubes are sheared. Figure 8(d) depicts an exemplified shear failure stress of ice cubes as a function of shearing distance for PA and PE substrates along y- and z-directions. It demonstrates however ice cubes show higher shear failure stress along the z-direction on PE surface, but shear failure stresses are almost the same along y-direction on PA and PE surfaces in which both substrates are atomistically smooth as shown in Figure 8(c). Figure 8(b) indicates shear failure stress of ice cubes as a function of interfacial displacements that are obtained by assigning x=0 in the linear fitting functions obtained in figure 5(a) in the model section. It shows at the same interfacial displacement, ice cube requires higher shear forces on PE compared to PA substrates that result in higher interfacial shear failure stresses on PE substrates as shown in Figure 8(a). We conclude that the surface lattice texture structure is a major dominant factor in determining ice shear stress in this study.

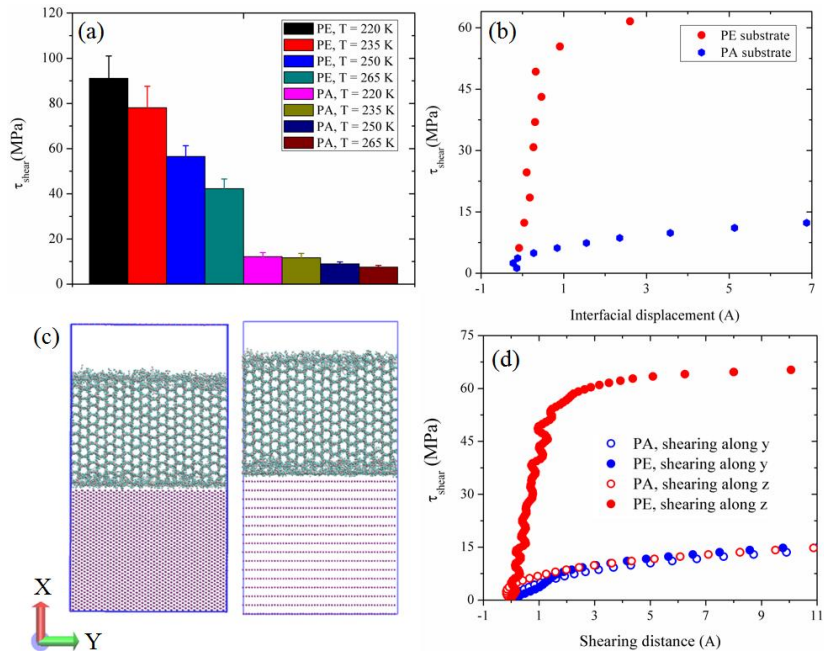

**Fig. 8.** (a) Shear failure stress along the z-direction for the PA and PE with nanoscale roughness of 0.335nm obtained at different four temperatures of 220K, 235K, 250K, and 265K. (b) Shear failure stress of ice cube as a function of interfacial displacement for PE (red dots) and PA (blue dots) substrates; (c) Snapshots of equilibrated ice-substrate systems along XY plane and (d) Shear failure stress of ice cube vs. shearing distance on PA and PE graphite substrates along y-(blue dots) and z-directions (red dots).

## D. EFFECT OF DEPTH OF WATER INTERDIGITATED ON ICE SHEAR STRENGTH MEASURMENT

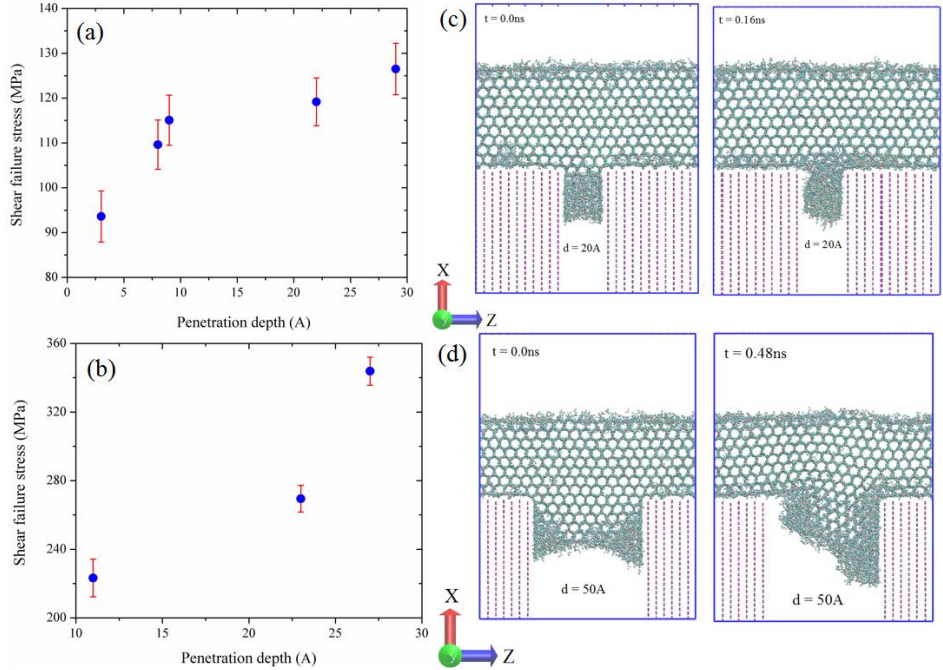

**FIG. 9.** Shear failure stress of ice as a function of the depth of water penetration for the PE samples with the surface nanoscale roughness size $d$ of (a) 2nm and (b) 5nm. Snapshots are taken during the shear-mode simulations of an ice cube on PE substrates with surface confinement size d of (c) 2nm and (d) 5nm, along the z-direction.

The effect of penetration depth (PD) of interlocked water molecules on the measured interfacial shear failure stress are depicted in Figure 9 (a)-(b) for the PE substrates with nanoscale roughness of d=2 and 5nm. Results indicate systems demonstrating higher PD result in increasing interfacial shear failure stress in which the rate of increase for the case of d=5nm, shown in Figure 9(b), is remarkably higher than that for the system with d=2nm. We address this difference to the distinction in strain distribution within the ice cubes on samples with different nanoscale roughness as follows. This difference is more obvious based on the snapshots of systems of ice cubes experiencing shear simulation along z-direction provided in Figure 9(c)-(d). For the system with nanoscale roughness of d=2nm, the interlocked water molecules penetrating the groove structure between the two graphite blocks are in a disordered supercooled state. Upon applying the shear load along the z-direction, the strain is centralized near the penetrating layer before failure occurs (the picture on the right of Figure 9(c)). Compare with the samples with larger nanoscale roughness of d=5nm, it is an obvious big portion of interlocked penetrating water molecules is in crystalline structure as shown in snapshots in Figure 9(d). When shear loads are applied, noticeable structural deformation is induced far away from the penetrating layer well into the ice on top of the substrate which is responsible for the stronger dependence on PD observed in Figure 9(b).

## E. EFFECT OF SURFACE ROUGHNESS ON ICE SHEAR FAILURE STRESS

Figure 10 shows interfacial shear failure stress of ice cube as a function of surface nanoscale roughness sizes of *d* value for the PA and PE cases for the systems indicate PD of about 1~1.6nm (except the PA and PE samples with *d* of 0.335nm). It indicates similar magnitudes of the shear failure stress along the y- and the z-directions obtained for the PA samples as to the atomistically smoothness of the PA substrates along with both principle directions. While the shear failure stress is more sensitive to the direction of loaded shear in PE substrates being six times greater along the z-direction than that along the y-direction. This is indeed the reflection of the "ruggedness" of the arrangements of graphene sheets in graphite blocks of PE cases along the z-direction at the ice-substrate interface, as shown in Figure 4 (b) and (c). The "ruggedness" is, however, absent along the y-direction, depicted in Figure 8(c) and as a result, our simulations find that the failure stress measured along the y-direction is indeed similar in magnitude to that of the PA samples (red symbols in Figure 10). Furthermore, it is understandable that shear failure stress exhibits a monotonic linear-like increase with increasing nanoscale roughness of *d* values for the systems being sheared along z-direction (the blue dots). The increase can be understood as the result of shear failure switching from being an interfacial (on PA and PE with d=0.335nm) to a cohesive mode with PE substrate having larger *d* values. The cohesive shear failure mode demands deterioration of the internal structure of the ice layer. It is expected that the cohesive failure mode of ice in PE substrates will become the dominant mode as *d* increases, and hence give rise to higher measured failure stresses.

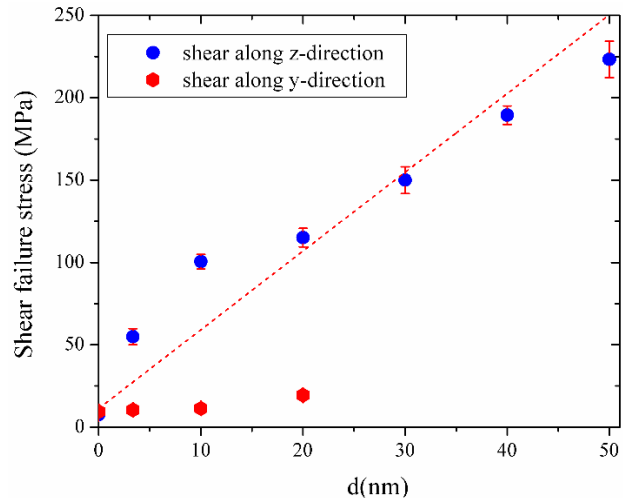

**FIG. 10.** The shear failure stress along the z-direction (blue symbols) and y-direction (red symbols) as the function of surface roughness size *d*. In this figure, PA and PE samples are reported as d = 0 nm and 0.335nm, respectively.

## F. ICE SHEARING DYNAMICS

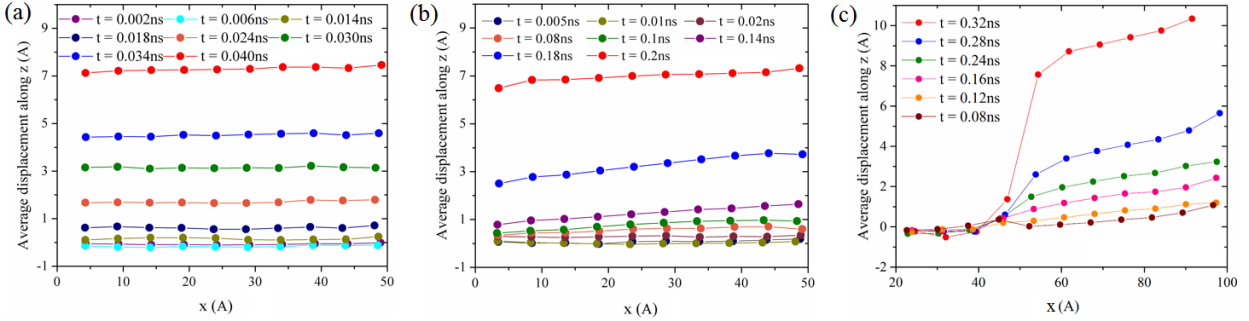

**FIG. 11.** Average displacement of ice cube as a function of x-direction calculated at different simulation time during shear simulations along with z-direction for the (a) PA, (b), and (c) PE substrates with nanoscale roughness size of 0.335nm and 2.0nm.

To investigate the shearing dynamics of an ice cube on different PA and PE graphite substrates, we performed an in-depth analysis of the shearing displacement of ice layers during the simulation time along the 'x' direction, as shown in Fig. 11. For this, we divide ice cube into several same-sized bins along x-direction perpendicular to the direction of loading shear force and calculate the average displacement of all water molecules included in each bin. Results show a significant change in displacement of different layers of ice cube along z-direction once shear failure stress is reached. Comparing to the displacement profile of ice cubes on PA substrate, the average displacement of different layers of ice cubes is increased on PE surfaces at a fixed simulation time, which could be addressed to the higher interfacial shear stress on PE cases. Furthermore, ice cube on PE substrate with the nanoscale roughness of d=20Å shows zero displacements for the 20 Å thickness of water molecules interlocked between graphene sheets, which is agreeable by substrate confinement illustrated in snapshots provided in Figure 4.

## IV. Conclusion

In this study, we use all-atom molecular dynamics simulation to examine nanoscale ice shearing strengths on smooth and texture graphite substrates consisting of different nanotexture surface sizes to investigate the shearing mechanics of ice at the atomistic scale. It is found the corrugation of the surface substrate, ice temperature, size of the surface roughness and the depth of water interdigitated between surface texture structures mainly influence the interfacial shear strength of ice. Compared to the previous MD simulation of ice fracture, our results demonstrate the effect of surface lattice structure on examining the ice shear strength in which rougher surface results in higher interfacial shear stress. We find that the higher system temperature results in increasing the mobility of interfacial water molecules and at the same time decreasing ice shear strength on various graphite substrates. It is understood that the shear strength of ice is extremely sensitive to the corrugation structure of the surface in which ice shear stress is about six times higher along z-direction compared to the y-direction that the surface texture is absent. Furthermore, our results show, for the first time, the effect of depth of interdigitated water molecules on interfacial ice shear stress. It is shown the shear failure stress is non-linearly increased with incrementing the depth of interlocked water molecules between graphene sheets. Our simulation results supply atomistic attributes of ice shearing consequences which improve our understanding of experimental achievements.

**DATA AVAILABILITY**

The data that support the findings of this study are also openly available in our Zenodo Repository at https://doi.org/10.5281/zenodo.4285668.

**ACKNOWLEDGEMENT**

A. Afshar and D. Meng acknowledge the support of NASA Glenn Research Center through the Advanced Aircraft Icing Subproject of the Advanced Air Transport Technology (AATT) Project (Cooperative Agreement NNX16AN20A, Richard E. Kreeger Technical Monitor). The support of the MSU Center for Advanced Vehicular Systems is also gratefully acknowledged.